\documentclass[twocolumn,superscriptaddress,showpacs,pra]{revtex4}
\usepackage{epsfig}
\usepackage{epstopdf}
\usepackage{delarray}
\usepackage{amsmath, amssymb}
\usepackage{bm}
\usepackage{graphicx}
\usepackage{placeins}
\usepackage{color}

\begin{document}
\title{Early Detection of Rogue Waves Using Compressive Sensing}

\author{Cihan Bayindir}
\email{cihan.bayindir@isikun.edu.tr}
\affiliation{Engineering Faculty, I\c{s}\i k University, \.{I}stanbul, Turkey}

\begin{abstract}
We discuss the possible usage of the compressive sampling for the early detection of rogue waves. One of the promising techniques for the early detection of the oceanic rogue waves is to measure the triangular Fourier spectra which begin to appear at the early stages of their development. For the early detection of the rogue waves it is possible to treat such a spectrum as a sparse signal since we would mainly be interested in the high amplitude triangular region located at the central wavenumber. Therefore compressive sampling can be a very efficient tool for the rogue wave early warning systems. Compressed measurements can be acquired by remote sensing techniques such as coherent SAR which measure the ocean surface envelope or by insitu techniques such as spectra measuring tools mounted on a ship hull or bottom mounted pressure gauges. By employing a numerical approach we show that triangular Fourier spectra can be sensed by compressive measurements at the early stages of the development of rogue waves such as those in the form of Peregrine and Akhmediev-Peregrine solitons. Our results may lead to development of the early warning hardware systems which use the compressive sampling thus the memory requirements for those systems can be greatly reduced.

\end{abstract}
\maketitle


\section{Introduction}
Rogue waves are commonly defined as the waves with a height more than 2-2.2 times the significant wave height in a wave field \cite{Kharif}. They present a danger to life, marine travel and operations but they are desired in optical fibers for communication purposes. Their result can be catastrophic and costly in marine environment \cite{Kharif, bayindir2015}. The early detection of rogue waves in the chaotic ocean is a must to ensure the safety of the marine travel and the offshore structures in stormy conditions \cite{bayindir2015}. This vital problem has been disregarded for a long time and has only been studied for almost a decade \cite{bayindir2015}. However early detection of rogue waves is an extremely hard and complicated problem due to many processes involved \cite{bayindir2015}. First of all rogue waves appear in stormy conditions therefore accurate prediction of weather conditions is a must. Secondly rogue waves appear in a length (time) scale that are on the order of their width \cite{Akhmediev2011}. Due to the rapidly changing nature of the rogue waves, the reliability of the forecast of the rogue waves currently is not very high and it is hard to expect that it will become more reliable in the near future \cite{Akhmediev2011}. One of the promising approaches which address this problem is to continuously measure the part of the whole surface spectrum in real time and use the triangular Fourier spectra of the growing rogue waves in early stages of their development in a chaotic wave field before the dangerous peak appears \cite{Akhmediev2011}.

In this paper, we discuss the possible usage of the compressive sampling  for the early detection of the oceanic rogue waves. We use a similar methodology to the one introduced in \cite{Akhmediev2011}. Similarly we analyze the emerging triangular rogue wave spectra, however we obtain those spectra via compressed measurements with a significant undersampling ratio. In order to sense the triangular spectra via compressive sampling, we offer a procedure as follows: in the time evolving field we take randomly selected undersampled measurements from the ocean surface fluctuations. Then we reconstruct the sparse triangular rogue wave spectra via compressive sampling. Then we move to the next time step and repeat the same procedure. By implementation of a numerical method we show that using compressive sampling technique we can detect the emergence of the triangular rogue wave spectra using significantly less samples compared to the classical sensing techniques. This method can directly be used for the early detection of the analytical rogue waves in optical fibers and wave flumes. However the usability of the method for the chaotic wave environments such as the realistic open ocean presents many difficulties and precisely how the compressive measurements could be accomplished in practice is beyond the scope of this letter. It is possible to use remote sensing techniques such as coherent SAR \cite{bay2013, bayindir2015jnse} or insitu techniques such as a device installed on a ship which measure the spectrum of only a part of the whole surface quickly by scanning the water surface bit-by-bit \cite{Akhmediev2011}. However as shown in this paper the compressive sampling can significantly reduce the memory requirements and the cost of rogue wave early warning and detection systems. 

\section{Review of the NLSE and its Rogue Wave Solutions}

\noindent Dynamics of weakly nonlinear deep water ocean waves can be described by the nonlinear Schr\"{o}dinger equation (NLSE) \cite{Zakharov1968}. It has been previously shown that the NLSE can also be used as a model to describe oceanic rogue waves \cite{Akhmediev2009b, Akhmediev2009a, Akhmediev2011, bayindir2015}. One of the most widely used forms of the nondimensional NLSE is given by
\begin{equation}
i\psi_t + \frac{1}{2} \psi_{xx} +  \left|\psi \right|^2 \psi =0
\label{eq01}
\end{equation}
where $x,t$ are the spatial and temporal variables, $i$ denotes the imaginary number and $\psi$ is complex amplitude \cite{bayindir2015}. This notation is mainly used in ocean wave theory however the $t$ and $x$ axes are switched in the optics studies \cite{bayindir2015}.  NLSE is also widely used in other branches of the applied sciences and engineering to describe various phenomena including but not limited to Bose-Einstein condensates, pulse propagation in optical fibers and quantum state of a physical system. Integrability of the NLSE is studied extensively within last forty years and many exact solutions of the NLSE are derived. Some rational soliton solutions of the NLSE are derived as well. One of the most early forms of the rational soliton solution of the NLSE is the Peregrine soliton which is considered as an accurate rogue wave model \cite{Akhmediev2009b, Peregrine}. It is given by
\begin{equation}
\psi_1=\left[1-4\frac{1+2it}{1+4x^2+4t^2}  \right] \exp{[it]}
\label{eq02}
\end{equation}
where $t$ is the time and $x$ is the space parameter. It is shown that Peregrine soliton is a first order rational soliton solution of the NLSE and higher order rational solutions also exist \cite{Akhmediev2009b}. Additionally many simulations of the chaotic wavefields have revealed that rogue waves with height more that $3$ can not be described by the Peregrine soliton \cite{Akhmediev2011, bayindir2015}. Therefore second and higher order rational soliton solutions are used to describe the rogue waves with heights bigger than $3$. It has been shown that second order rational soliton solution of NLSE, with a peak height $5$, can be a model for the oceanic rogue waves as well  \cite{Akhmediev2011, bayindir2015}.  It has also been confirmed that the second order rational soliton solution of the NLSE still presumes a triangular Fourier spectra  \cite{Akhmediev2011, bayindir2015}. The second order rogue wave which satisfies the NLSE exactly is known as Akhmediev-Peregrine soliton is given as \cite{Akhmediev2009b}
\begin{equation}
\psi_2=\left[1+\frac{G_2+it H_2}{D_2}  \right] \exp{[it]}
\label{eq03}
\end{equation}
where
\begin{equation}
G_2=\frac{3}{8}-3x^2-2x^4-9t^2-10t^4-12x^2t^2
\label{eq04}
\end{equation}
\begin{equation}
H_2=\frac{15}{4}+6x^2-4x^4-2t^2-4t^4-8x^2t^2
\label{eq05}
\end{equation}
and
\begin{equation}
\begin{split}
D_2=\frac{1}{8} [ \frac{3}{4} &+9x^2+4x^4+\frac{16}{3}x^6+33t^2 \\
& +36t^4+\frac{16}{3}t^6-24x^2t^2+16x^4t^2+16x^2t^4 ]
\end{split}
\label{eq06}
\end{equation}
Even higher order rational solutions of the NLSE and a hierarchy of obtaining those rational solutions based on Darboux transformations \cite{Matveev} are given in \cite{Akhmediev2009b}. Additionally, throughout many simulations it has been confirmed that rogue waves obtained by numerical techniques which solve the NLSE are in the forms of these first (Peregrine) and higher order rational solutions of the NLSE \cite{Akhmediev2009b, Akhmediev2009a, Akhmediev2011}. One of the promising techniques for the early detection of the rogue waves is to use the triangular Fourier spectra that begins to develop at the early stages of the rogue waves \cite{Akhmediev2011b} therefore the theoretical shapes of the Fourier spectra of the rogue waves need to be discussed. The Fourier transform of the Peregrine soliton can analytically be calculated as 
\begin{equation}
F(k,t)=\frac{1}{\sqrt{2 \pi}}\int_{-\infty}^{\infty} \psi(t,x)e^ {ikx}dx
\label{eq07}
\end{equation}
which gives
\begin{equation}
\begin{split}
F(k,t)=\sqrt{2 \pi}[\frac{1+2it}{\sqrt{1+4t^2}}\exp{\left(-\frac{\left|k\right|}{\sqrt{2}}\sqrt{1+4t^2}\right)} &-\delta(k)] \\
& .\exp{[it]}
\end{split}
\label{eq08}
\end{equation}
where $k$ is the wavenumber parameter and $\delta$ is the Dirac-delta function \cite{Akhmediev2011}. The Fourier spectra of the first, second and higher order rogue waves are compared and discussed in \cite{Akhmediev2009b} in detail where some analytical expressions and mainly numerical and illustrative results are presented. Similar to the first order rogue wave, the second order rogue wave has roughly a triangular Fourier spectrum \cite{Akhmediev2011b} when the dirac delta peak due to constant term is ignored. However compared to the Fourier spectrum of the first order rogue wave, the Fourier spectrum of the second order rogue wave exhibits two dips due to increased number of sidebands in the wave profile \cite{bayindir2015, Akhmediev2011b}.

Although a numerical solution is not necessary for the analytical solutions presented above, considering future research that involves many spectral components we solve the NLSE using a split-step Fourier method (SSFM) which is one of the widely used Fourier spectral methods with efficient time integration. SSFM performs the time integration by time stepping of the exponential function for an equation which includes a first order time derivative \cite{bayindir2015}. In SSFM, like other spectral techniques \cite{bay2015c, bay2015d, trefethen}, the spatial derivatives are calculated by the orthogonal transforms  \cite{bayindir2015arxivcsmww, bayindir2015arxivcssfm, bay2015e}.  Some of their applications can be seen in \cite{bayindir2015, bay2009, Demiray2015, Karjadi2010, Karjadi2012, bayindir2015d, bayindir2015arxivwvlt, bayindir2015arxivchbloc, bayindir2016KEE, bay2015b, bayAnExtnd, bayNoisyTun, bayCSRM, bayZenoRogue} and more detailed discussions can be seen in \cite{trefethen}. In the SSFM, the time integration is performed by time stepping of the exponential function for an equation which includes a first order time derivative \cite{bayindir2015}. SSFM is based on the idea of splitting the governing equation into two parts, the nonlinear and the linear part. For the NLSE, the advance in time due to nonlinear part can be written as \cite{bayindir2015}
\begin{equation}
i\psi_t= -\left| \psi \right|^2\psi
\label{eq11}
\end{equation}
which can be exactly solved as
\begin{equation}
\tilde{\psi}(x,t_0+\Delta t)=e^{i\left| \psi(x,t_0)\right|^2\Delta t}\ \psi(x,t_0)
\label{eq12}
\end{equation}
where $\Delta t$ is the time step. Taking linear part of the NLSE as \cite{bayindir2015}
\begin{equation}
i\psi_t=-\frac{1}{2}\psi_{xx}
\label{eq13}
\end{equation}
Using the Fourier series one can write that
 \begin{equation}
\psi(x,t_0+\Delta t)=F^{-1} \left[e^{-ik^2\Delta t/2}F[\tilde{\psi}(x,t_0+\Delta t) ] \right]
\label{eq14}
\end{equation}
where $k$ is the Fourier transform parameter \cite{bayindir2015}. Therefore combining the expressions in (\ref{eq12}) and (\ref{eq14}), the complete formulation of the SSFM can be written as
 \begin{equation}
\psi(x,t_0+\Delta t)=F^{-1} \left[e^{-ik^2\Delta t/2}F[ e^{i\left| \psi(x,t_0) \right|^2\Delta t}\ \psi(x,t_0) ] \right]
\label{eq15}
\end{equation}
which is used to calculate the surface fluctuations, $\psi$, starting from the initial conditions. We start the rogue wave simulations using the analytical rogue wave solutions mentioned above. The number of spectral components are selected as $N=4096$ in order to make use of the FFTs efficiently. The time step is selected as $dt=0.05$ which does not cause any stability problems. The actual water surface fluctuation for this initial condition would be given by the real part of $ \left|\psi\right| \exp{[i\omega t]}$ where $\omega$ is some carrier wave frequency.  However in the present study we only consider the envelope, $ \left|\psi\right|$.

\section{Review of the Compressive Sampling}

After it has been introduced to the scientific community with a seminal paper by \cite{Candes2006}, compressive sampling (CS) has become a core research area in the last decade. In summary, CS states that a sparse signal can be reconstructed from fewer samples than the samples that Nyquist-Shannon sampling theorem states. Currently it is a common tool in various branches of applied mathematics and engineering and currently many software and hardware systems make use of this efficient signal processing technique. In this section we try to sketch a brief summary of the CS.

Let $\eta$ be a $K$-sparse signal of length $N$, that is only $K$ out of $N$ elements of the signal are nonzero. $\eta$ can be represented using a orthonormal basis functions with transformation matrix ${\bf \lambda}$. Typical transformation used in literature are Fourier, discrete cosine or wavelet transforms just to mention few. Therefore one can write $\eta= {\bf \lambda} \widehat{ \eta}$ where $\widehat{ \eta}$ is the transformation coefficient vector. Since $\eta$ is a $K$-sparse signal one can discard the zero coefficients and obtain $\eta_s= {\bf \lambda}\widehat{ \eta}_s$  where $\eta_s$ is the signal with non-zero elements only.

The idea underlying in the CS is that a $K$-sparse signal $\eta$ of length $N$ can exactly be reconstructed from $M \geq C \mu^2 ({\bf \Phi},{\bf \lambda}) K \textnormal{ log (N/K)}$ measurements with an overwhelmingly high probability, where $C$ is a positive constant and $\mu^2 (\Phi, \lambda)$ is coherence between the sensing basis ${\bf \Phi}$ and transform basis ${\bf \lambda}$ \cite{Candes2006}. Taking $M$ random projections by using the sensing matrix ${\bf \Phi}$ one can write  $g={\bf \Phi} \eta$. Therefore the problem can be recognized as
\begin{equation}
\textnormal{ min} \left\| \widehat{ \eta} \right\|_{l_1}   \ \ \ \  \textnormal{under constraint}  \ \ \ \ g={\bf \Phi} {\bf \lambda} \widehat{ \eta}
\label{eq09}
\end{equation}
where $\left\| \widehat{ \eta} \right\|_{l_1}=\sum_i \left| \widehat{ \eta}_i\right|$. So that among all signals which satisfies the given constraints, the ${l_1}$ solution of the CS problem can be given as  $\eta_{{}_{CS}} ={\bf \lambda} \widehat{ \eta}$. $l_1 $ minimization is only one of the alternatives which can be used for obtaining the solution of this optimization problem. There are some other algorithms to recover the sparse solutions such as greedy algorithms \cite{candes2006compressive}. A more detailed discussion of the CS can be seen in \cite{Candes2006}.
It is useful to note that we are using the sparsity property of the Fourier transform of $\psi$. Therefore we can write ${\bf \lambda} \psi=\eta$ where $\eta$ is sparse triangular spectra, ${\bf \lambda} $ is the Fourier transformation matrix and $\psi= \widehat{ \eta}$ is the sparse surface fluctuation envelope measurement.

\section{Compressive Sampling of the Triangular Rogue Wave Spectra}

As discussed in \cite{Akhmediev2011}, the triangular shape of the Fourier spectra can be used for the early detection of the rogue waves since it becomes evident at the early stages of their development.  This is true both for the first and the second order rational solitons and also validated for the rogue waves in a chaotic wave field  \cite{Akhmediev2011, bayindir2015}.

\begin{figure}[htb!]
\begin{center}
   \includegraphics[width=3.4in]{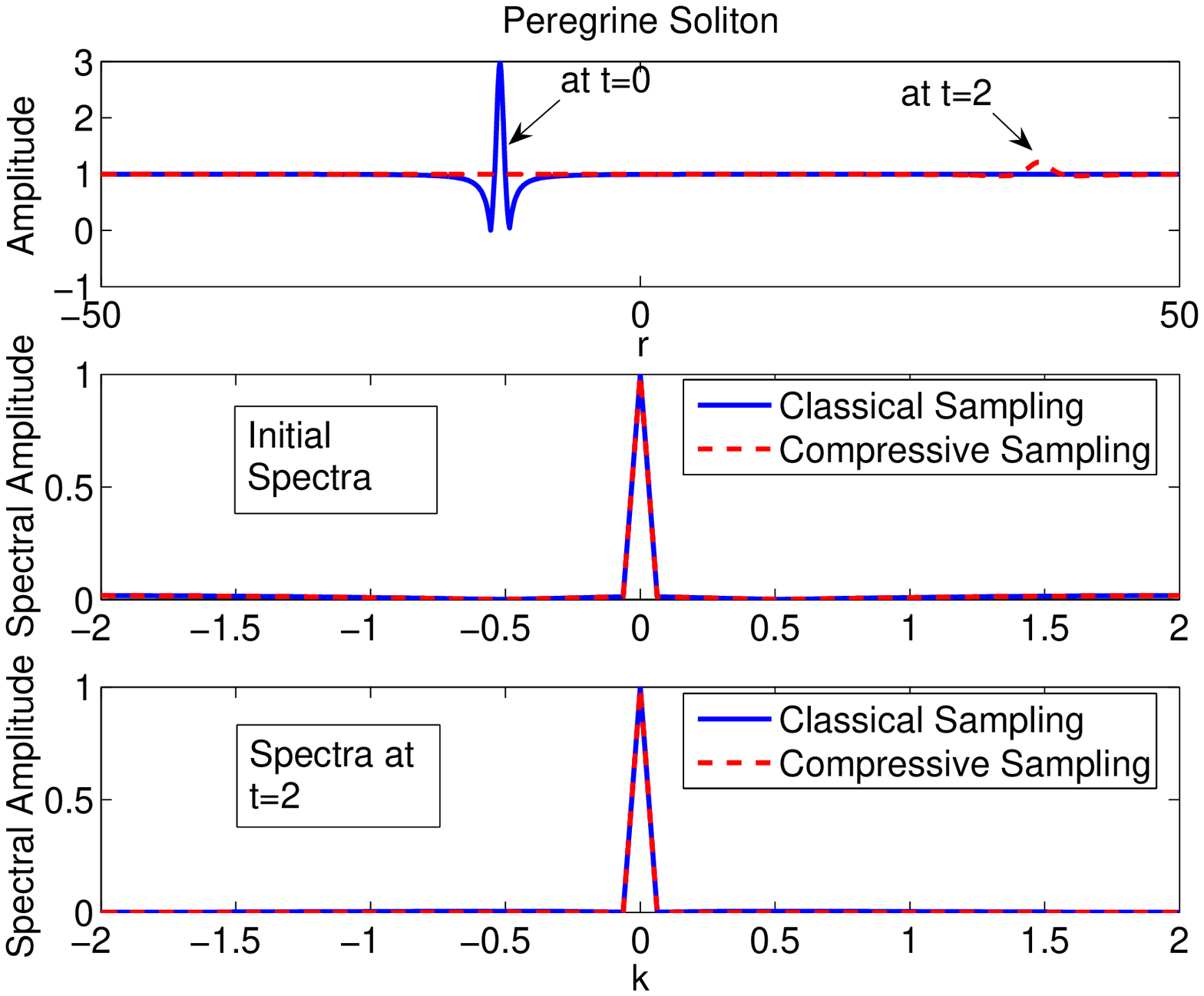}
  \end{center}
\caption{\small a) Peregrine soliton at $t=0$ and $t=2$ b) the Fourier spectrum of the Peregrine soliton at $t=0$ obtained by $N=1024$ classical and $M=64$ compressive samples c) the Fourier spectrum of the Peregrine soliton at $t=2$ obtained by $N=1024$ classical and $M=64$ compressive samples}
  \label{fig1}
\end{figure}

 The results for the Peregrine soliton are depicted in Fig.~\ref{fig1} where first subfigure shows the 1D Peregrine soliton in the physical domain at $t=0$ and $t=2$ which can be time reversed symmetrically considering the early detection purposes. The second and the third subfigures show the spectra obtained at $t=0$ and $t=2$, respectively. In this figure the continuous line refers to the spectra obtained classical $N=1024$ components whereas the dashed line refers to same spectra obtained using random $M=64$ compressive samples.
The normalized root-mean-square (rms) difference of the two spectra presented in the second subfigure obtained by the classical and compressive sampling techniques at time $t=0$ is $1.55\times10^{-10}$ whereas the rms difference between two results presented in third subfigure becomes  $7.91\times10^{-4}$ at time $t=2$. In our simulations we observe that CS can be used to obtain the sparse spectra more efficiently by employing fewer number of compressive samples when the rogue wave is at its peak.

\begin{figure}[htb!]
\begin{center}
   \includegraphics[width=3.4in]{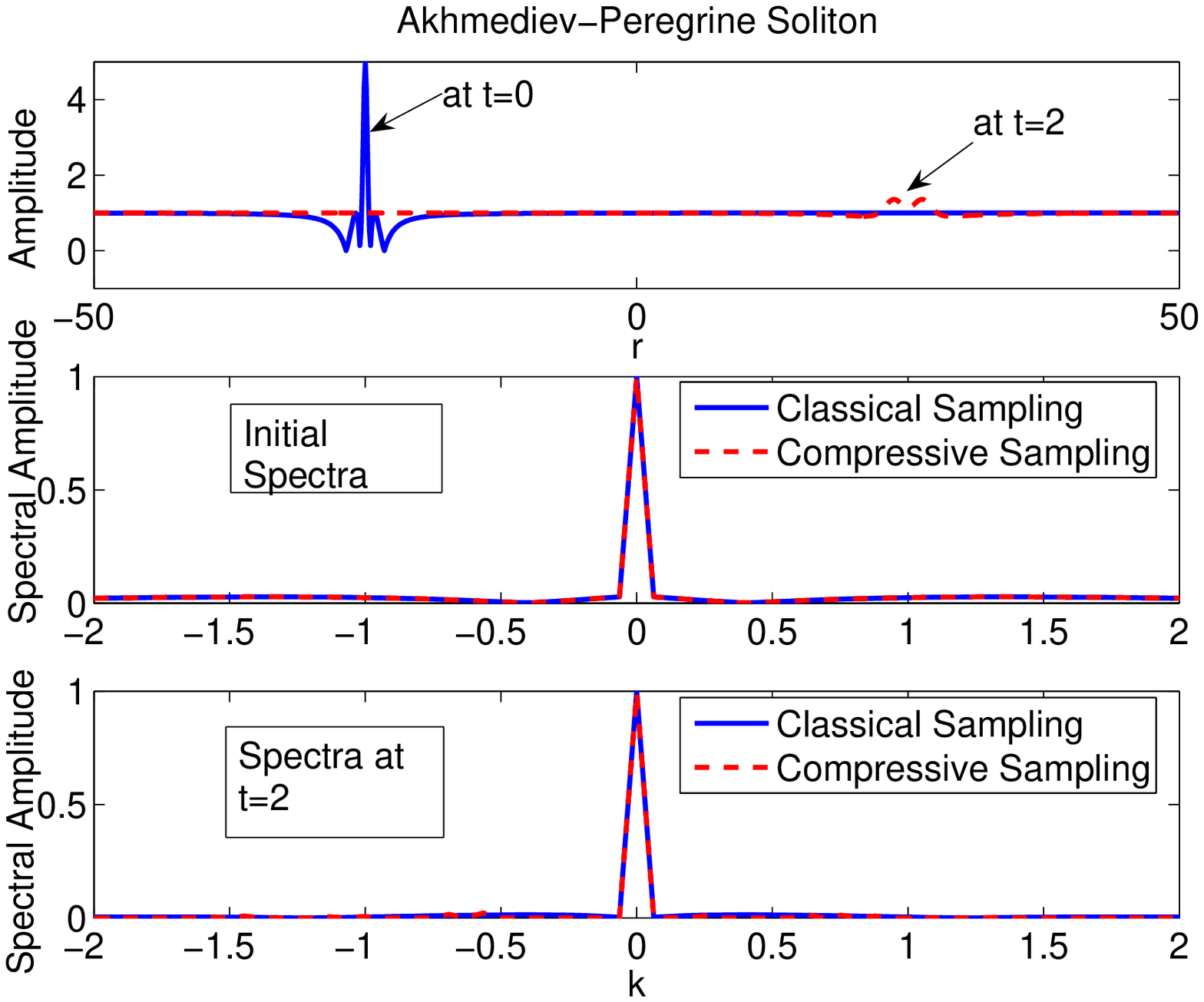}
  \end{center}
\caption{\small a) Akhmediev-Peregrine soliton at $t=0$ and $t=2$ b) the Fourier spectrum of the Peregrine soliton at $t=0$ obtained using $N=1024$ classical and $M=64$ compressive samples c) the Fourier spectrum of the Peregrine soliton at $t=2$ obtained using $N=1024$ classical and $M=64$ compressive samples.}
  \label{fig2}
\end{figure}

	The results for Akhmediev-Peregrine soliton are depicted in Fig.~\ref{fig2} where first subfigure shows the 1D Akhmediev-Peregrine soliton in the physical domain at $t=0$ and $t=2$ which can be time reversed symmetrically considering the early detection purposes. The second and the third subfigures show the spectra obtained at $t=0$ and $t=2$, respectively. In this figure again the continuous line refers to the spectra obtained classical $N=1024$ components whereas the dashed line refers to same spectra obtained using random $M=64$ compressive samples.
The normalized root-mean-square (rms) difference of the two spectra presented in the second subfigure obtained by the classical and compressive sampling techniques at time $t=0$ is $1.60\times10^{-3}$ whereas the rms difference between two results presented in third subfigure is $2.70\times10^{-3}$ at time $t=2$. In our simulations we observe that CS can be used to obtain the sparse spectra more efficiently by employing fewer number of compressive samples when the Akhmediev-Peregrine soliton is at its peak.

The great advantage of the CS based methodology proposed in this paper is obvious. We can detect the developing Fourier spectra of the emerging rogue waves using CS with significantly less samples compared to the classical sensing theory. This result may be immediately applied in fiber optics and hydrodynamic wave flume studied. It may also lead to development of low cost remote and insitu sensing devices with significantly less memory requirements compared to the classical sensing devices. It also can transform the rogue wave early warning technology, that is, rather than directly measuring the spectra it may become easier and more efficient to measure the time series with compressed measurements which can be adapted to measurement systems currently in use. Currently we can not answer how the hardware implementation would be done since it requires many optimizations such as selecting the area for measurement, adjusting the noise and sensitivity of the electronic equipment, estimating the effects of complications due to 2D structure, nonlinearity and dispersion of waves. However the results presented in this paper are promising to enhance the rogue wave early detection technology.

\section{Conclusion}
In this paper we have discussed the possible usage of the compressive sampling for the early detection of rogue waves emerging in a chaotic sea state. One of the promising techniques for the early detection of rogue waves is to measure the triangular Fourier spectra which begin to appear at the early stages of the development of the oceanic rogue waves. Recognizing that such spectra can be treated
as a sparse signal, since we would mainly be interested in the central high amplitude triangular region for the early detection purposes of the rogue waves, the compressive sampling technique can become an efficient tool for the rogue wave early warning systems. 
Employing a chaotic wave field simulation we have showed that emerging triangular rogue wave spectra can be detected using the compressive sampling technique from significantly less samples compared to the classical sensing methods. These measurements can either be acquired by remote sensing techniques such as coherent SAR or insitu techniques such as spectra measuring tools mounted on a ship hull or bottom mounted pressure gauges.  Our results show that the compressive sampling based methodology proposed in this paper can reduce the memory requirements of the early warning hardware systems significantly therefore their efficiency can be enhanced while their cost is reduced.


\begin{thebibliography}{00}

\bibitem{Kharif}
C. Kharif and E. Pelinovsky. European Journal of Mechanics, B: Fluids.  {\bf{6}}, 603 (2003).

\bibitem{bayindir2015}
C. Bay\i nd\i r. Physics Letters A,  {\bf{380}}, 156 (2016).

\bibitem{Akhmediev2011}
N. Akhmediev, J. M. Soto-Crespo, A. Ankiewicz and N. Devine. Physics Letters A,  {\bf{375}}, 2999 (2011).

\bibitem{bay2013}
C. Bay\i nd\i r. PhD Thesis, Georgia Institute of Technology (2013).

\bibitem{bayindir2015jnse}
C. Bay\i nd\i r. Journal of Naval Science and Engineering,  {\bf{11}}, 68 (2015).

\bibitem{Zakharov1968}
V. E. Zakharov. Soviet Physics JETP,  {\bf{2}}, 190 (1968).

\bibitem{Akhmediev2009b}
N. Akhmediev, A. Ankiewicz and J. M. Soto-Crespo. Physics Review E,  {\bf{80}}, 026601 (2009).

\bibitem{Akhmediev2009a}
N. Akhmediev, J. M. Soto-Crespo and A. Ankiewicz. Physics Letters A,  {\bf{373}}, 2137 (2009).
		
\bibitem{Peregrine}
D. H. Peregrine. Journal of Australian Mathematical Society: Series B,  {\bf{25}}, 16 (1983).

\bibitem{Matveev} V. B. Matveev and M. A. Salle.  Springer-Verlag. Berlin (1991).

\bibitem{Akhmediev2011b}
N. Akhmediev, A. Ankiewicz, J. M. Soto-Crespo and J. M. Dudley. Physics Letters A,  {\bf{375}}, 541 (2011).

\bibitem{bay2015c}  C. Bay\i nd\i r. Hesaplamal\i \ ak\i\c{s}kanlar mekani\u{g}i  \c{c}al\i\c{s}malar\i \ i\c{c}in s\i k\i\c{s}t\i r\i labilir Fourier tayf\i \ y\"{o}ntemi, XIX. T\"{u}rk Mekanik Kongresi, Trabzon, (2015). (In Turkish)

\bibitem{bay2015d}  C. Bay\i nd\i r. Okyanus dalgalar\i n\i n s\i k\i \c{s}t\i r\i labilir Fourier tayf\i \ y\"{o}ntemiyle h\i zl\i  \ modellenmesi, XIX. T\"{u}rk Mekanik Kongresi, Trabzon, (2015). (In Turkish)

\bibitem{trefethen}
L. N. Trefethen. Spectral Methods in {MATLAB}, (2000).

\bibitem{bay2015e}  C. Bay\i nd\i r. S\"{o}n\"{u}ml\"{u} de\u{g}i\c{s}tirilmi\c{s} Korteweg de-Vries (KdV) denkleminin analitik ve hesaplamal\i \ \c{c}\"{o}z\"{u}m kar\c{s}\i la\c{s}t\i rmas\i, T\"{u}rk Mekanik Kongresi, Trabzon, (2015). (In Turkish)

\bibitem{bayindir2015arxivcssfm}
C. Bay\i nd\i r. Compressive Split-Step Fourier Method. arXiv Preprint,  arXiv:1512.03932 (2015).

\bibitem{bayindir2015arxivcsmww}
C. Bay\i nd\i r. Compressive spectral method for the simulation of the water waves. arXiv Preprint,  arXiv:1512.06286 (2015).

\bibitem{bay2009}
C. Bay\i nd\i r. MS Thesis, University of Delaware (2009).

\bibitem{Demiray2015}
H. Demiray and C. Bay\i nd\i r. Physics of Plasmas,  {\bf{22}}, 092105 (2015).

\bibitem{Karjadi2010}
E. A. Karjadi, M. Badiey and J. T. Kirby. The Journal of the Acoustical Society of America,  {\bf{127}}, 1787 (2010).

\bibitem{Karjadi2012}
E. A. Karjadi, M. Badiey, J. T. Kirby and C. Bay\i nd\i r. IEEE Journal of Oceanic Engineering,  {\bf{37-1}}, 112 (2012).

\bibitem{bayindir2015d}
C. Bay\i nd\i r. TWMS: Journal of Applied and Engineering Mathematics,  {\bf{5-2}}, 298 (2015).

\bibitem{bayindir2015arxivwvlt}
C. Bay\i nd\i r. Early Detection of Rogue Waves by the Wavelet Transforms. arXiv Preprint,  arXiv:1512.02583 (2015).

\bibitem{bayindir2015arxivchbloc}
C. Bay\i nd\i r. Shapes and Statistics of the Rogue Waves Generated by Chaotic Ocean Current. arXiv Preprint,  arXiv:1512.03584 (2015).

\bibitem{bayindir2016KEE}
C. Bay\i nd\i r. Rogue waves of the Kundu-Eckhaus equation in a chaotic wave field. arXiv Preprint,  arXiv:1601.00209 (2016).

\bibitem{bay2015b}  C. Bay\i nd\i r. Analytical and numerical aspects of the dissipative nonlinear Schr\"{o}dinger equation. TWMS: Journal of Applied and Engineering Mathematics,  vol. 6, no.1, 135 (2016).

\bibitem{bayAnExtnd}  C. Bay\i nd\i r. An extended Kundu-Eckhaus equation for modeling dynamics of rogue waves in a chaotic wave-current field. arXiv Preprint, arXiv:1602.05339 (2016).

\bibitem{bayNoisyTun}  C. Bay\i nd\i r. Rogue wavefunctions due to noisy quantum tunneling potentials. arXiv Preprint, arXiv:1604.06604 (2016).

\bibitem{bayCSRM}  C. Bay\i nd\i r. Compressive spectral renormalization method. arXiv Preprint, arXiv:1611.08551 (2016).

\bibitem{bayZenoRogue}  C. Bay\i nd\i r and F. Ozaydin. Freezing rogue waves by quantum Zeno dynamics. arXiv Preprint, arXiv:1701.01997(2017).

\bibitem{Candes2006} Candes, E. J., Romberg, J. and Tao, T. IEEE Transactions on Information Theory, 52, 489-509, (2006).

\bibitem{candes2006compressive} 
E. J. Candes. Proceedings of the international congress of mathematicians, {\bf{3}}, 1433, (2006).
















 \end{thebibliography}
\end{document}